\RequirePackage{fix-cm}
\documentclass[smallextended]{svjour3}       
\smartqed  
\usepackage{graphicx}
\usepackage{amsmath}
\usepackage{amssymb}
\usepackage{upgreek}
\usepackage{xcolor}
\usepackage{cite}
\usepackage{lmodern}

\begin{document}

\title{Upconversion based receivers for quantum hacking resistant quantum key distribution
}

\author{Nitin Jain         \and
        Gregory S. Kanter 
}
\institute{N. Jain \and  G.S. Kanter \at
Department of Electrical Engineering and Computer Science,\\
2145 Sheridan Road, Northwestern University, Evanston IL 60208-3118 \\
              \email{nitin.jain@northwestern.edu}           
}

\date{Received: date / Accepted: date}

\maketitle

\begin{abstract} 
We propose a novel upconversion (sum frequency generation) based quantum-optical setup that can be employed as a receiver (Bob) in practical quantum key distribution systems. The pump governing the upconversion process is produced and utilized inside the physical receiver, making its access or control unrealistic for an external adversary (Eve). This pump facilitates several properties which permit Bob to define and control the modes that can participate in the quantum measurement. Furthermore, by manipulating and monitoring the characteristics of the pump pulses, Bob can detect a wide range of quantum hacking attacks launched by Eve. 
\end{abstract}

\section{Introduction}
The process of upconversion of light denotes a nonlinear interaction in which two optical frequencies $\omega_1$ and $\omega_2$ add together to generate optical radiation at a third frequency $\omega_3 = \omega_1 + \omega_2$. It was first observed in lithium niobate almost half a century ago~\cite{Midwinter1967}. A quantum-mechanical treatment~\cite{Tucker1969,Kumar1990} of such frequency conversion processes suggested that the quantum features of one of the input light beams, usually called `signal', could be preserved during the nonlinear interaction. This has now been confirmed in numerous experiments~\cite{Huang1992,Kim2001,Vandevender2004,Diamanti2005,MaL2012} dealing with frequency upconversion of quantum-optical states, e.g. entangled states, weak coherent states. 

A majority of upconversion experiments aim to convert optical signals in the near-infrared regime to those in the visible or near-visible regime, where highly-efficient and low-noise detectors are easily available. Improvements in upconversion devices~\cite{Thew2006,Pelc2011,Pelc2012} and novel approaches, such as the quantum pulse gate~\cite{Eckstein2011} and mode resolving detection enabled by optical arbitrary waveform generation~\cite{Huang2013,Kowligy2014}, offer the possibility of high-dimensional communication with large throughputs. With help of photonic quantum information interfaces~\cite{Tanzilli2005}, the prospects of realistic quantum communication networks~\cite{Gisin2007,Kimble2008} become more promising. 

Quantum key distribution (QKD) is certainly one area that can benefit from the aforementioned advances. Briefly, QKD is a cryptographic method that facilitates two users Alice (sender) and Bob (receiver) to obtain symmetric keys that can then be used for encrypting messages in an information-theoretically secure manner~\cite{BB84,Gisin2002,Scarani2009a}. An adversary Eve attempting to eavesdrop on the key exchange induces errors that can be detected by Alice and Bob, thus alerting them and preventing the loss of confidentiality of their encrypted messages. 

In the last decades, QKD has evolved to become the foremost application of quantum information technology. Upconversion technology has already been successfully demonstrated for a variety of QKD systems~\cite{Diamanti2005,MaL2012}, including implementations of differential phase shift~\cite{Inoue2002,Zhang2009} and measurement-device-independent (MDI) protocols~\cite{Lo2012,Liu2013}. With features such as low jitter~\cite{Thew2006} and inherent polarization sensitivity~\cite{Xu2007} in addition to the improved detection efficiency (via Si avalanche photodiodes), upconversion-assisted detectors have been instrumental in pushing up the secret key rates in practical QKD.

The theoretical security model of QKD, which fundamentally utilizes the laws of quantum mechanics, offers provably secure key exchange. In practice however, imperfections and limitations of the physical hardware can open security loopholes that can be exploited by Eve to attack the QKD system and obtain knowledge of the key without alerting Alice and Bob~\cite{Scarani2009,Scarani2009a,Lo2014}. 
Such attacks are popularly referred to as quantum hacking~\cite{VadimWebsite}. 

A majority of quantum hacking attempts demonstrated in the last years involved Eve actively injecting optical pulses into the QKD receiver to exploit imperfections such as timing related issues in the detection apparatus. The intensity of these pulses (sometimes accompanied with CW power ranging from less than $1~$mW to a few Watts) ranged from that of dim coherent states containing $10-100$s of photons to bright pulses having millions of photons~\cite{Wiechers2011,Lydersen2010a,Gerhardt2011,Sauge2011,Jain2011,LydersenJain2011,Weier2011,Bugge2014}. In most cases, while the wavelength of Eve's light was the same as that used by Alice and Bob, i.e. the telecom wavelength of $\sim 1550\,$nm, a few eavesdropping strategies have also been explored at different wavelengths~\cite{Li2011,Jain2014,Jain2015}. 	

In this paper, we propose an upconversion protected (UCP) receiver that would fortify practical QKD against a wide range of quantum hacking attacks~\cite{Kanter2015}. The design of the UCP-QKD receiver employs off-the-shelf optical and optoelectronic components and its operation is simple. The design also obviates the usual difficulties in trying to analyze (the security of) a QKD system where every component is evaluated over a huge range of variables such as wavelength, time, etc. The resulting boost in security improves the overall performance of a non-cryogenically cooled QKD system since higher quality semiconductor (Silicon) based single photon detectors are employed. Therefore, even though we are of the opinion that no cryptographic system -- quantum or classical -- can be guaranteed to be unconditionally secure, the proposed receiver has some desirable properties that enhance the security and performance of practical QKD. In that sense, our work significantly adds to the advantages of upconversion technology for practical QKD systems. It also joins the growing body of countermeasures~\cite{Silva2012,Lucamarini2015} and novel schemes, such as MDI-QKD~\cite{Lo2012}, conceived in order to heal the Achilles' heel of QKD~\cite{AchillesHeelQKD}: the (vulnerable) photon detection system, without compromising on performance.

The organization of the paper is as follows: in section~\ref{ucdfQKDrxr}, we present the basic principles on which the security provided by the UCP-QKD receiver is based. We also propose a prototype, explaining its design, operation, and the implementation of the security features in detail. In section~\ref{protcn} thereafter, we discuss how the prototyped receiver provides protection against three classes of attacks. After discussing some practical aspects in section~\ref{ovprac} to showcase that the performance of the UCP-QKD receiver is comparable to state-of-the-art QKD implementations, we conclude in section~\ref{conc}.
\section{Upconversion based QKD receivers\label{ucdfQKDrxr}}
\begin{figure}[!t]
\centering
\includegraphics[width=11.5cm]{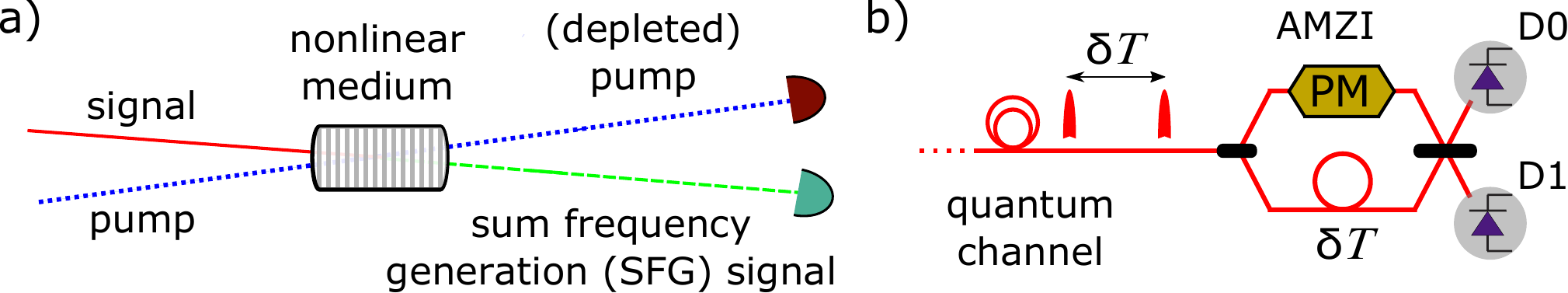}
\caption{(Color online) Basic upconversion device and quantum key distribution receiver schemes. a) The upconversion takes place inside the nonlinear medium when the pump and signal modes overlap as prescribed by the phase-matching conditions. The resulting upconverted/SFG photons inherit the quantum properties of the signal. b) Generic QKD receiver used by Bob to interferometrically detect quantum states coded in phase and time by Alice and sent over the quantum channel. AMZI: asymmetric Mach Zehnder interferometer, PM: phase modulator, D0/1: single photon detectors. \label{upconvNqkdrcvr}}
\end{figure}

Figure~\ref{upconvNqkdrcvr}a shows the main idea of upconversion. A strong classical `pump' field at wavelength $\lambda_{\rm pump}$ interacts with a `signal' field at wavelength $\lambda_{\rm sig}$ inside a $\chi^{(2)}$ medium, e.g. a nonlinear waveguide, for the sum frequency generation (SFG) process. The `upconverted/sum' radiation is at the wavelength $\lambda_{\rm sum} = \left(\frac{1}{\lambda_{\rm pump}} + \frac{1}{\lambda_{\rm sig}}\right)^{-1}$ and can be shown to preserve the quantum statistics of the signal field. This is due to the fact that the field operators of the signal and upconverted modes simply transform into each other (neglecting an overall phase factor) at an appropriate pump amplitude~\cite{Tucker1969,Kumar1990,Huang1992}. 

In order to implement quantum communication systems with upconversion devices, two useful figures of merit are the efficiency $\eta \equiv \eta_c \eta_d$ and the (dark) noise level in the measurement of the SFG light. 
Here, $\eta_c$ essentially indicates the probability to \emph{convert} the wavelength of a signal photon from $\lambda_{\rm sig}$ to $\lambda_{\rm sum}$ in the nonlinear process, while $\eta_d$ is the probability to then \emph{detect} that single photon. Note that any practically-useful definition of $\eta_c$ would include the waveguide's non-unity transmission, due to coupling and propagation losses in the nonlinear medium. The noise includes the intrinsic dark counts of the detector and those due to parasitic nonlinear interactions, such as pump-induced parametric fluorescence and spontaneous Raman scattering~\cite{Dong2008,Pelc2011}. Based on the typical values of $\eta_c$ and $\eta_d$ reported in numerous experiments, achieving $\eta \sim 40\%$ at a dark count probability below $10^{-6}$ per signal pulse should be feasible with commercially available detectors~\cite{Pelc2011,MaL2012}. Receivers utilizing upconversion technology can therefore surpass the performance of those detecting photons directly at telecom wavelengths (using non-cryogenic detectors)~\cite{Itzler2011}.
\subsection{Building blocks \label{ucdfQKDrxr:bb}}
Figure~\ref{upconvNqkdrcvr}b shows a generic time-multiplexed interferometric scheme~\cite{Bennett92c} that is at the heart of several practical QKD receiver setups (both prepare-and-measure and entangled-based type)~\cite{Hughes2000,Ekert1992,Marcikic2004,Yuan2005a}. The photon is in a superposition of the two optical modes, temporally separated by $\delta T$, that form the basis of the qubit space and are used for encoding of the bits ``0'' and ``1''. This quantum signal, sent over the quantum channel, is received and decoded in an asymmetric Mach Zehnder interferometer (AMZI) that consists of a phase modulator (PM) and a pair of single photon detectors (D0 and D1). The PM is used for active basis selection in prepare-and-measure systems and for adjusting the interferometer to maintain the correlations between pairs of detectors across the receivers of Alice and Bob in entangled-based systems. Unless mentioned explicitly, we will be referring to prepare-and-measure systems only. 

Integration of the upconversion device shown in Fig.~\ref{upconvNqkdrcvr}a in such a QKD receiver requires the quantum states received over the quantum channel (assumed to be in control of Eve) to be upconverted by the pump before the detection stage. Guaranteeing secure operation then implies that the detection and basis selection apparatus in Bob must be isolated from the external influences of Eve. More specifically, Eve's injection from the channel into the receiver must be restricted in wavelength, time, and power. In the following, we describe several properties -- intrinsic to upconversion devices -- that facilitate achieving such an objective. The building blocks of the secure UCP-QKD receiver are thus based on the proper implementation of these properties.
\subsubsection*{Linear wavelength isolation (LWI)}
An upconversion system can be designed so that no optical wavelength linearly propagates through the system. In other words, only light generated by a specified local nonlinear interaction will reach the detection apparatus in the receiver. Such linear wavelength isolation (LWI) can be achieved simply by using optical filters, such as a band pass filter (BPF) which lets through light at wavelengths only inside a certain band. As an instance, although Eve can inject as much power as desired at the signal wavelength, without the SFG interaction, an ideal BPF with central wavelength $\lambda_{\rm sum}$ would block the injected light from reaching the detectors\footnote{Security concerns arising from nonlinear interactions in the waveguide, especially the conversion of photons injected by Eve at the signal wavelength after interacting with Bob's pump, will be examined in the next subsections.}. With a carefully-chosen width, the BPF will also be effective in blocking the harmonics ($\lambda_{\rm sig}/n$, where $n\geq 2$ is an integer) that may get generated due to poorly phase-matched nonlinearities. A short pass filter (SPF) that cuts off any light at wavelengths above a specified wavelength could also be employed, for example to block Eve's radiation at $\lambda_{\rm pump}$ from the quantum channel. This prevents Eve from controlling the nonlinear conversion process herself. Thus, only the frequencies converted via a specified nonlinear interaction eventually reach the detector while all others are suppressed. The detectors are then effectively isolated -- in the linear regime -- at \emph{all} optical wavelengths injected at the input of the receiver. (In general, by using an appropriate combination of optical filters before and after the waveguide, any leakage can also be ensured to be negligible). The LWI property thus heavily curtails the wavelength degree of freedom for Eve.  
\subsubsection*{Optical gating (OG)}
In the so-called gated mode operation of a detector, e.g. single photon avalanche diode (SPAD), an electronically-applied gate raises the diode's sensitivity to obtain a measurable response to even single photons in a short temporal window. In an upconversion device, the nonlinear interaction inside the waveguide can happen only in the presence of the pump pulse. This allows us to correspondingly define a gate of width $\tau_{\rm gate}$ where detection events can be recorded, thereby restricting Eve's exploitation via the temporal degrees of freedom. 
\begin{figure}[!t]
\centering
\includegraphics[width=11.0cm]{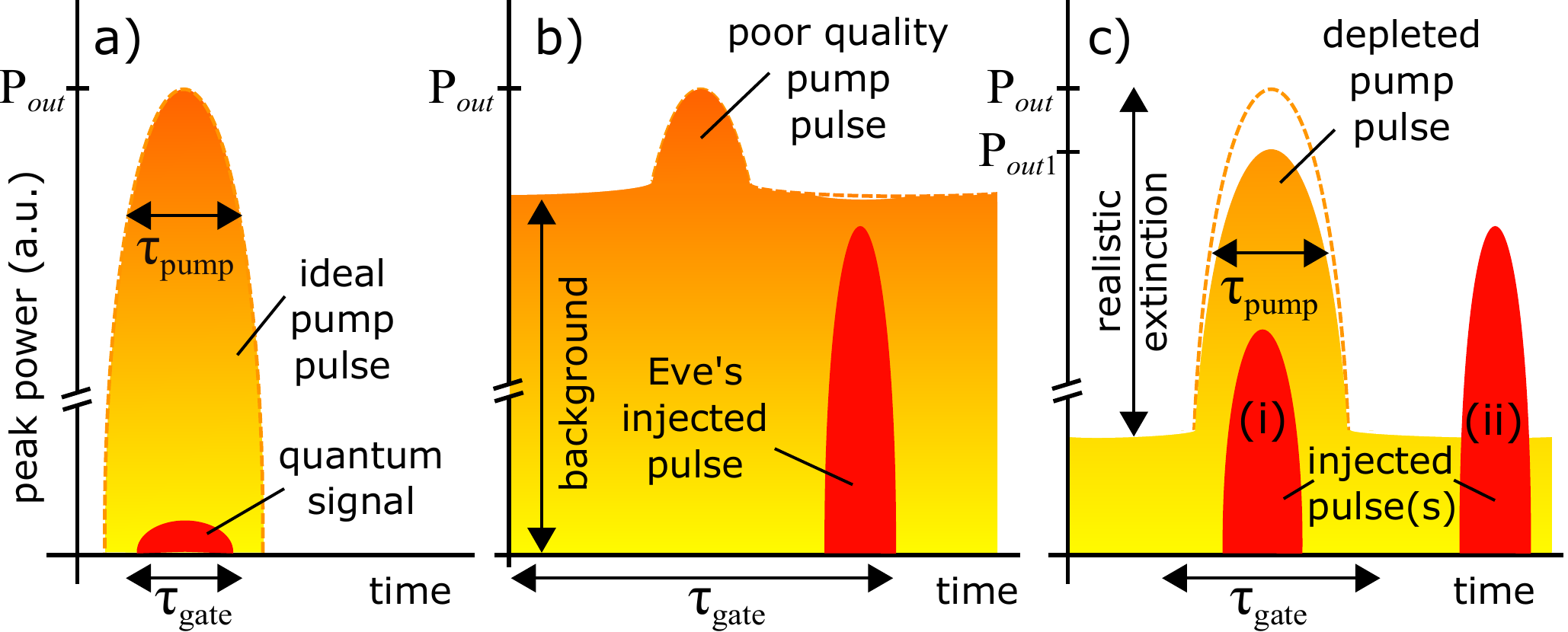}
\caption{(Color online) Illustration of the power limiting and monitoring (PLM) and optical gating (OG) properties. (Assume the y-axis with a logarithmic scale). The pump pulse defines an optical gate where upconversion can happen. a) For an ideal pulse, $\tau_{\rm gate} \approx \tau_{\rm pump}$ holds. With just the quantum signal, the undepleted (dashed border) and depleted pump pulses coincide. b) In case of poor extinction however, the pump-pulse specified gate is either ill-defined or too wide; $\tau_{\rm gate} >> \tau_{\rm pump}$. Eve may then escape the monitoring system by shifting the strong pulse to upconvert pump background photons. c) Many of Eve's attacks in which she replaces the quantum signals with strong pulses, for instance, those labeled (i) or (ii), can be defeated if the pump pulses can be produced with reasonable extinction ($\tau_{\rm gate} \gtrsim \tau_{\rm pump}$) and their depletion can be monitored with high resolution.\label{fPLMnOG}}
\end{figure}
However, the definition of $\tau_{\rm gate}$ is strongly dependent on the extinction ratio $ER$ achieved in the production of the pump pulses. In an ideal case with infinite $ER$, as depicted in Fig.~\ref{fPLMnOG}a, it is clear that only signals having an overlap with the pump pulse would undergo the nonlinear conversion. At the other extreme, a poor $ER$ (translating into a relatively-large background level, as shown in Fig.~\ref{fPLMnOG}b) combined with pulse jitter can force $\tau_{\rm gate}$ to be much wider than $\tau_{\rm pump}$. This may be exploited by Eve, e.g., through bright faked states arriving slightly shifted away from the pump peak, as we shall explain later. The extinction ratio must therefore be large enough to exclude the possibility of these type of attacks, as depicted by pulse (ii) in Fig.~\ref{fPLMnOG}c. Finally, the pump in fact may also act as a gate in the wavelength regime if the frequency conversion process is phase-matched only for one signal wavelength.
\subsubsection*{Power limiting and monitoring (PLM)}
The amount of optical power entering any QKD receiver can both be limited and monitored, e.g. by using appropriate combinations of passive and active optical components. Recently however, vulnerabilities in such countermeasures were reported~\cite{Jain2015}. 
For an upconversion device, the limiting function is available in a fundamental sense: the photon flux of the upconverted pulses is upper bound by the photon flux of the pump pulses. Moreover, a powerful pulse in the signal mode, as shown by pulse (i) in Fig.~\ref{fPLMnOG}c, would lead to a depletion of the pump at the output of the waveguide. To elaborate, let us denote the peak pump power measured at the input [output] of the waveguide by $P_{in}$ [$P_{out}$] in the absence of the signal (so that the ratio $P_{out} / P_{in}$ indicates the true optical loss through the waveguide). Bob could then monitor the actual power $P_{out1}$ in the presence of a signal. Any noticeable depletion of the pump would indicate a bright pulse with multiple photons -- likely from Eve's injection. In a quantitative sense, any unusual fluctuations of the quantity $\zeta = |1-P_{out1}/P_{out}|$ could be used for raising alarm and alerting Alice. Remarkably, this monitoring provides a \emph{zero insertion loss method} to locate signatures of potential attacks. The pump parameters and the resolution of the monitoring system are critical to the success of the power limiting and monitoring (PLM) property. Note that it is also possible to losslessly monitor light at the signal wavelength, which further enhances the ability of Bob to discover unwanted signals injected by Eve. 
\subsubsection*{Nonlinear application of basis choice (NABC)}
The SFG signal has a phase that is the combination of both the signal and pump phases~\cite{Boyd2008}. This opens up the possibility of selecting the basis for the signal in a nonlinear fashion by randomly modulating the pump phase during the upconversion process. Notably, this kind of basis selection can be performed actively and is essentially loss-less for the signal. A straightforward security advantage of NABC is that Eve cannot access this nonlinear phase shift. Furthermore, selecting the basis using the pump \emph{alone} implies the isolation of Bob's basis selection apparatus from Eve without using any special components such as optical isolators. This of course is subject to how well the pump path can be insulated from Eve. As we shall describe later in detail, the chances of accessing the secret basis choice remotely from the quantum channel by means of sending bright pulses and analysing back-reflections can be made negligible~\cite{Gisin2006,Jain2014}. Additionally Bob can use the NABC property to detect the presence of injected pulses from Eve, such as faked states. 
\subsection{System design and operation \label{sysdsgnnop}}
The design and operation of a generic upconversion QKD receiver which is secure against a host of quantum hacking attacks is governed by the above properties. In a QKD system that implements these properties correctly, Bob can both define and control the modes that participate in the quantum measurement, which ultimately decides the exchanged key. 

\begin{figure}
\centering
\includegraphics[width=11.0cm]{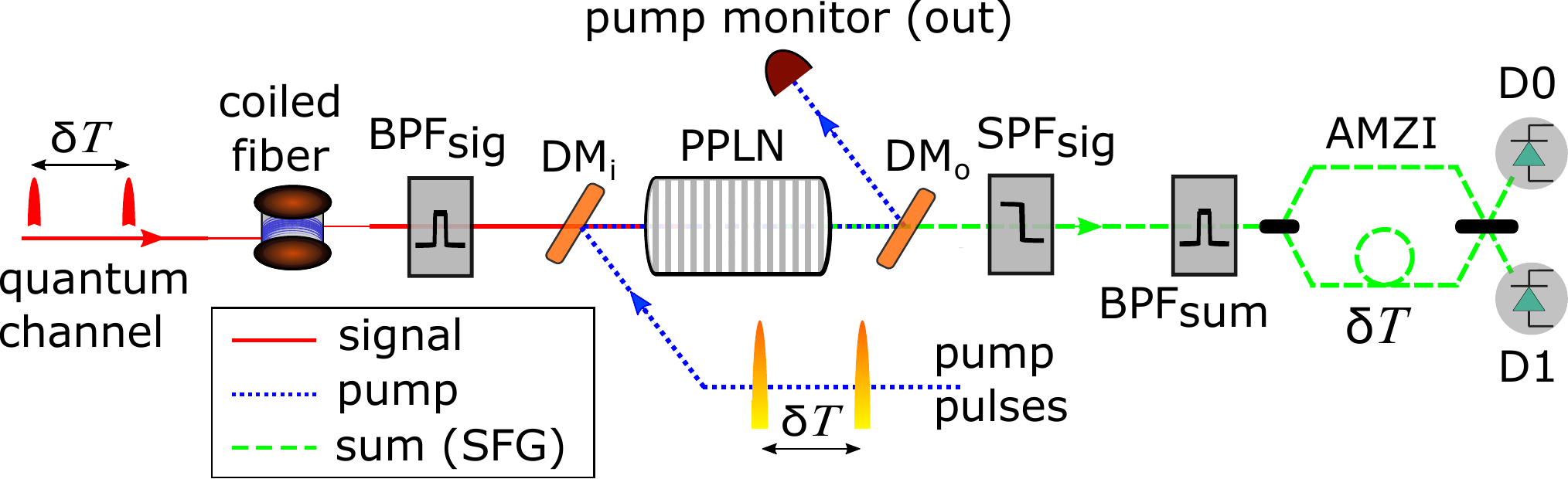}
\caption{(Color online) Schematic of the UCP-QKD receiver. The description and role of the various filters, such as $\rm BPF_{sig}$, is given in the text. The AMZI in the pump path is not shown for the sake of clarity. It will be discussed later; see Fig.~\ref{nabcNplm} for instance. Using a DM, the pump and signal paths (dotted and solid lines) are overlapped before the waveguide. Similarly, the SFG (dashed line) is separated from the other waves after the waveguide. The $\rm DM_{i[o]}$ at the input[output] of the waveguide must be chosen to be highly transmissive at the signal[sum] wavelength. After passing through another AMZI, the upconverted photons at $\lambda_{\rm sum}$ are detected by the single photon detectors D0 and D1 (dashed line). Although not shown, the pump power is monitored before the waveguide as well. AMZI: asymmetric Mach Zehnder interferometer, B/SPF: Band/Short pass filter, DM: dichroic mirror \label{fSysDiag}}
\end{figure}

Figure~\ref{fSysDiag} shows an exemplary schematic of the UCP-QKD receiver based on the time-multiplexed interferometric scheme discussed in Fig.~\ref{upconvNqkdrcvr}b. 
Below we describe the various components and their roles; unless mentioned otherwise, we shall always refer to the operation of phase-coded BB84 protocol~\cite{BB84}.

For the nonlinear material, we consider a periodically poled lithium niobate (PPLN) waveguide for this example. Pump pulses are generated using a pulsed laser source at wavelength $\lambda_{\rm pump} \approx 1810\,$nm; the choice of this long wavelength pump is guided by decreased noise in the upconversion process~\cite{Pelc2011}. Every cycle, a pair of pump pulses interact -- in the PPLN waveguide -- with a pair of signal pulses (wavelength: $\lambda_{\rm sig} \approx 1530\,$nm) arriving from the quantum channel. The pulses in a pair are separated by time $\delta T$ and to ensure a high efficiency of upconversion, the pump pulse width $\tau_{\rm pump}$ is chosen so that it encompasses the signal during the nonlinear interaction ($\tau_{\rm sig} < \tau_{\rm pump}$). To minimize transmission losses, anti-reflection coatings at $\lambda_{\rm sig} [\lambda_{\rm sum}]$ can minimize the in[out]-coupling losses of the waveguide.

The upconverted modes at $\lambda_{\rm sum} \approx 830\,$nm are subsequently decoded in the asymmetric Mach Zehnder interferometer (AMZI) by reversing the delay $\delta T$. It can be seen that the time windows/bins of detection are defined by the pump pulses: detection events in one of three different output time bins labeled \texttt{early}, \texttt{middle}, and \texttt{late} are obtained every cycle. The relative phase of the upconverted modes, which determines the interference outcomes in the \texttt{middle} time bin, is controlled by the transfer of the phase of the pump pulses onto signal pulses during upconversion. Therefore the AMZI just before the detectors includes only a time delay. The outcomes are recorded via the detection clicks of Si-based single photon avalanche diodes (SPADs) that exhibit a high quantum efficiency around $\lambda_{\rm sum}$ and operate in the free-running mode. 
\subsubsection*{Implementation of LWI}
The coiled fiber at the entrance of the QKD receiver shown in Fig.~\ref{fSysDiag} essentially serves as a short pass filter (SPF)\footnote{Alternatively, an appropriate filter can be used, but a coiled fiber offers a lower loss and cost-effective solution.}. The radius of the coil is chosen such that the SPF cutoff wavelength is $\lesssim 1800\,$nm. This would have a minimal impact on the signal photons at $\lambda_{\rm sig} \approx 1530\,$nm and simultaneously prevent Eve from injecting her own pump at long(er) wavelengths. Together with the PPLN waveguide, featuring a transparency window that starts around $350\,$nm, the pertinent wavelength range for any attack could then be roughly narrowed down to $300-1850\,$nm. 

Using a dichroic mirror (DM) before/after the PPLN (see Fig.~\ref{fSysDiag}) facilitates in combining/separating the pump from the signal/upconverted photons. Depending on the dichroic filter type, the practical implementation may work in the reflecting or transmitting mode~\cite{ThorlabsDMSumPumpSep,OmegaDMSumPumpSep}. Note that if an all-fiber-integrated assembly is employed~\cite{Tanzilli2012,Harris2014}, wavelength division multiplexers can be used instead of the dichroic mirrors. The short pass filter $\rm SPF_{sig}$ after the waveguide cuts off the infrared pump and signal wavelengths while allowing the upconverted photons to pass through to the detectors. Together with the two band pass filters $\rm BPF_{sig}$ and $\rm BPF_{sum}$, centered at $\lambda_{\rm sig}$ and $\lambda_{\rm sum}$, respectively, it plays an essential role in implementing LWI in the receiver. In other words, this assembly ensures that light at \emph{all} wavelengths injected from the quantum channel are suppressed before the SPDs except for the photons upconverted from $\lambda_{\rm sig}$ to $\lambda_{\rm sum}$ by Bob's pump pulses. 

\begin{figure}[!t]
\centering
\includegraphics[width=11.0cm]{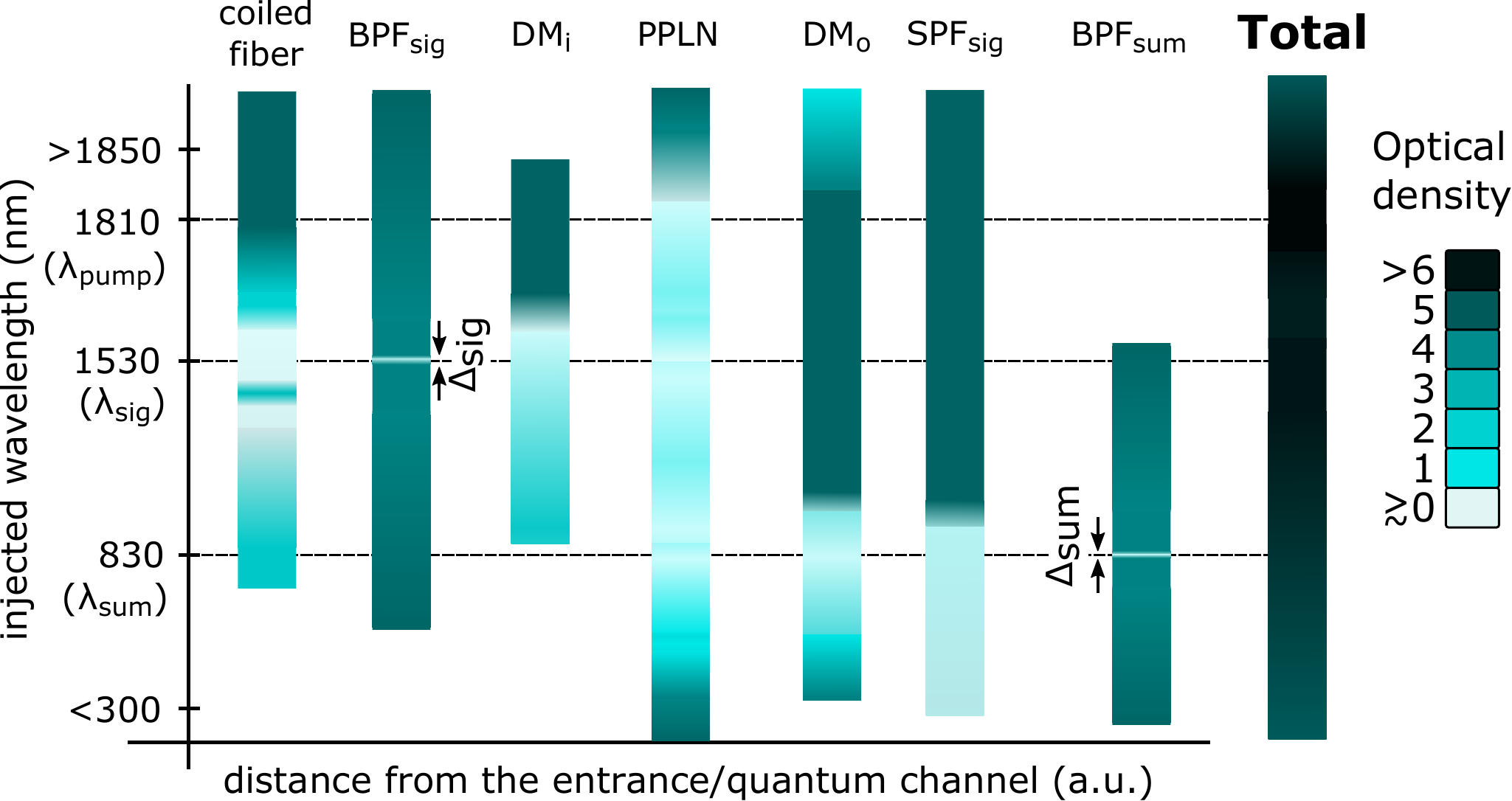}
\caption{(Color online) Linear wavelength isolation in the QKD receiver shown in Fig.~\ref{fSysDiag}. Any wavelength -- chosen by Eve to attack -- corresponds to a horizontal line in this chart. It is possible to engineer that any wavelength encounters heavy losses ($>60\,$dB) before reaching the SPDs. The gradient values for the different components here are illustrative. The widths $\Delta_{\rm sig}$ and $\Delta_{\rm sum}$ of the band pass filters $\rm BPF_{sig}$ and $\rm BPF_{sum}$ centered at $\lambda_{\rm sig}$ and $\lambda_{\rm sum}$ are not more than a few tens of nm; this choice does not affect the overall receiver efficiency.\label{fLWI}}
\end{figure}

Figure~\ref{fLWI} sums up the contribution of the different optical elements that realizes LWI in the receiver.
Several commercial DMs, SPFs, and BPFs that can be used here are available in both free-space and fiber-optic versions~\cite{opnetiWDM,Edmund85891,SemrockLL01830}. Also, two filters of the same kind could be cascaded to provide even higher optical density in the $300-1850\,$nm wavelength range. In fact, one could also get rid of the SPFs and just use a series of $\rm BPF_{sig}$ [$\rm BPF_{sum}$] before [after] the waveguide -- provided sufficient isolation in the aforementioned wavelength range (with the exception of $\lambda_{\rm sig}$ and $\lambda_{\rm sum}$) is achievable. 

It is very important that the net filtering greatly attenuate an input at the phase-matched pump wavelength. This ensures that Eve cannot inject a large pump -- together with signal of appropriate power level -- that allows her to have control over the nonlinear interaction and circumvent the power limiting and LWI properties of the system. Given the large wavelength separation between the pump and signal ($\lambda_{\rm pump} - \lambda_{\rm sig} \sim 300\,$nm), ultra-high pump attenuation with low signal loss should be possible. We estimate that $\sim 80\,$dB of isolation (which is quite easy to obtain because of the dichroic mirror) until the waveguide at $\lambda_{\rm pump}$ would be sufficient. This would ensure that the power of Eve's injected pump is $<100\,$nW inside the waveguide, and the corresponding upconverted light is too dim to be useful for any known attacks\footnote{We assume that the optical fibers in front of the receiver would readily fuse, disconnecting the receiver from the quantum channel, for $\gtrsim 10\,$W average input powers~\cite{Lucamarini2015}.}.
\subsubsection*{Implementation of PLM and OG}
To implement PLM, the residual pump (after the nonlinear interaction) is diverted to a monitoring stage where its depletion is recorded. A high resolution monitoring system would catch classical optical pulses injected by Eve in the signal mode, as was shown in Fig.~\ref{fPLMnOG}c. If $\zeta_{\rm min}$ denotes the minimum resolution of the monitoring system and $P_{out}$ denotes the pump peak power at the output of the waveguide in \emph{normal} circumstances, peak powers as low as $\zeta_{\rm min} P_{out} \lambda_{\rm pump} \tau_{\rm pump}/\lambda_{\rm sig} \tau_{\rm sig}$ of pulses substituted by Eve in the signal mode could be caught. Taking realistic pump and signal parameters of $P_{out} = 100\,$mW, $\tau_{\rm pump} = 100\,$ps and $\tau_{\rm sig} = 80\,$ps, and $\zeta_{\rm min}=0.1\%$, Eve's pulses with peak powers above $150\,\mu$W~\cite{Lydersen2010,Sauge2011,Wiechers2011,Gerhardt2011} would result in alarm and foil the attack. 
\begin{figure}[!t]
\centering
\includegraphics[width=11.0cm]{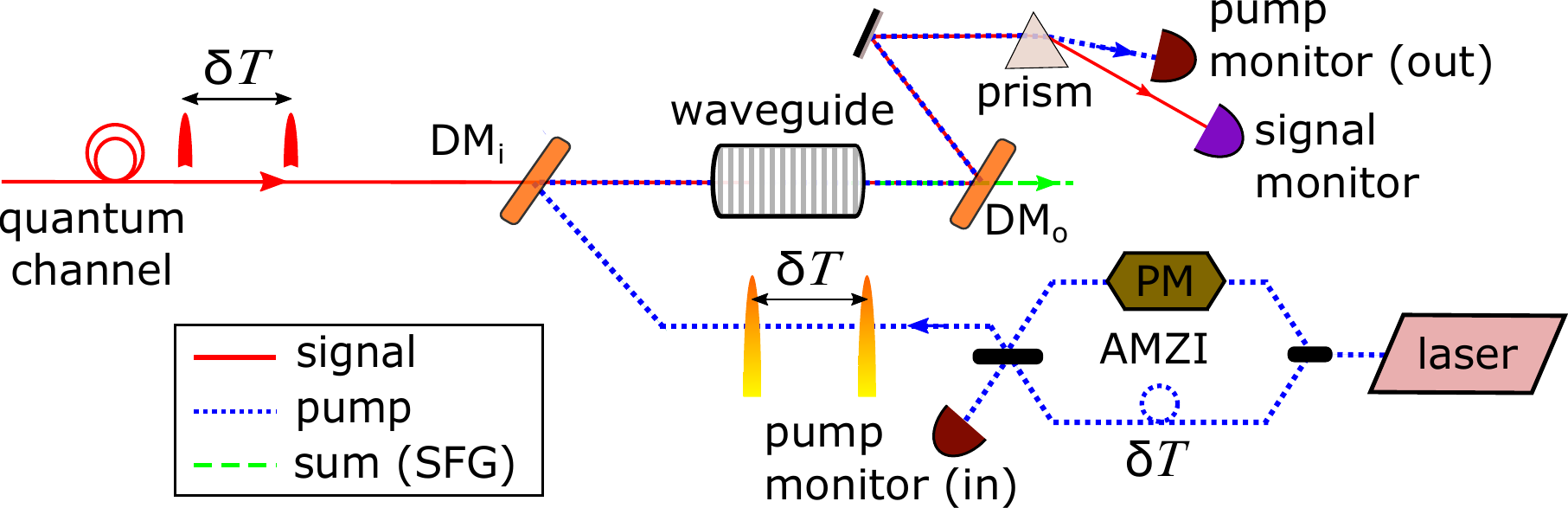}
\caption{(Color online) Monitoring of the pump \& signal and nonlinear phase transfer between them. Relative power measurements before and after the waveguide by pump monitors (in) and (out), respectively, can indicate even minor depletions. The depleted signal (solid red line after prism) can be optionally monitored with a suitably-sensitive detector that would register detection events during attacks. The relative phase modulated on the pump is transferred onto the signal during the nonlinear interaction. A prerequisite to implementing such basis selection is the calibration of the modulation voltages at $\lambda_{\rm pump}$ to ensure correct measurement of the QKD protocol alphabet with the upconverted photons.\label{nabcNplm}}
\end{figure}
Such high resolution monitoring can be performed by recording the pump power both before and after the waveguide, as illustrated in Fig.~\ref{nabcNplm}. 
The pulse monitoring could actually be taken one step further by also monitoring light at the signal wavelength. This could be done by adding a prism in the pump monitoring arm, as shown in Fig.~\ref{nabcNplm}. Nothing would be registered on this monitor in case of a legitimate quantum signal, however, Eve's large injections around $\lambda_{\rm sig}$ would get caught due to insufficient depletion. Notably, this is again a loss-less method to check for attacks. 

In order to implement OG securely, we believe $ER$ values of at least $-30\,$dB would be required. Consider a situation similar to the one depicted in Fig.~\ref{fPLMnOG}c with $\tau_{\rm gate} \sim 2\tau_{\rm pump}$. Such a large gate width may be practically necessary to minimize the effect of jitter, etc. Reusing the numerical values of the example in the previous paragraph, a bright faked state timed to arrive just after the pump-pulse peak (like pulse (ii) in the figure, but just inside $\tau_{\rm gate}$) avails at the most $P_{out}/1000 \approx 100\,\mu$W of power from the pump. Even assuming full upconversion, or $\eta_c=100\%$, the peak power of the pulses generated at $\lambda_{\rm sum}$ and hitting Bob's SPDs cannot exceed $\lambda_{\rm sig}/\lambda_{\rm sum} \times 100 \approx 185\,\mu$W, which is again well below the typical peak powers used in faked-state attacks\footnote{In fact, using the formula for conversion efficiency~\cite{Tucker1969,Kumar1990}, the actual $\eta_c < 1\%$ and with propagation losses around $3\,$dB, the peak power on the detectors would be $<1\,\mu$W.}.

To reduce even further the chances of security breach from Eve's attack pulses outside the pump pulse, one can also randomly modulate the phase modulator (PM) \emph{between} pump pulses. This would defeat the detector control strategy due to randomization of the detection outcomes in D0 and D1. In that sense, the impact of a poor $ER$ can also be mitigated. 
\subsubsection*{Implementation of NABC}
Figure~\ref{nabcNplm} shows a simple scheme that demonstrates the preparation of the pump pulses by using an AMZI in the pump arm; also see Fig.~\ref{fSysDiag}. 
This scheme also illustrates the nonlinear application of basis choice (NABC), some aspects of which have already been covered. In particular, using a dichroic mirror to combine the pump and signal can be quite effective in isolating the pump path from the quantum channel. As explained previously, this successfully prevents Eve from obtaining back-reflections, e.g. via Trojan-horse attacks~\cite{Gisin2006,Jain2014,Jain2015}, that may divulge the secret basis choice. 

With a simple modification in the setup of Fig.~\ref{nabcNplm}, Bob can also detect the presence of Eve's faked states, which adds to the already-available security features due to PLM. For this, Bob creates a single pump pulse instead of a pair. Such a task can be accomplished by using an intensity modulator in one of the arms of the pump AMZI. The upconversion now takes place for only one of the two signal modes resulting in only two instead of three time bins. Moreover, the interference in the \texttt{middle} bin is stamped out. Due to this, Eve cannot control the detection outcomes using faked states. This technique can be used to randomly sample the quantum channel for faked state injection. 
\section{Protection against various quantum hacking attacks\label{protcn}}
In the following, we discuss in detail how the generic QKD receiver can foil three specific categories of attacks. At the end of this section, we also explain the advantages of using free-running single photon avalanche diodes (SPADs) over their gated counterparts for the purpose of upconverted signal detection. 
\subsection{Detector control attacks\label{protcn:dca}} 
The basic idea of detector control attacks is to exploit certain imperfections of the detection subsystem to control the measurement outcomes. In the last decade, numerous such attacks have been demonstrated on a variety of practical QKD receivers~\cite{Zhao2008,Lydersen2010,Lydersen2010a,Wiechers2011,Gerhardt2011,Jain2011,LydersenJain2011,Weier2011}. In the case of the UCP-QKD receiver however, the optical gating (OG) and linear wavelength isolation (LWI) properties ensure that detection events can take place only in the presence of a pump pulse. As explained in the previous section, this is subject to a reasonable extinction of the pump pulses. Assuming the pump is well isolated from Eve's control, temporally advancing or delaying of Alice's states in the quantum channel~\cite{Zhao2008} or timing dim coherent faked states to prevent incorrect detection events~\cite{Jain2011} would necessarily result in a misalignment w.r.t. the pump, causing a severe reduction of the detection click rate. Note that the pump-defined temporal detection window creates a common gate for all subsequent detectors and monitors. Attacks that use blinding light with faked-state pulses~\cite{Lydersen2010,Lydersen2010a,Gerhardt2011} that have a temporal overlap with the pump pulse would be caught as they would lead to an observable depletion of the pump pulse; see Fig.~\ref{fPLMnOG}c as an example.

Using a single pump pulse instead of pulse pairs at random sample times, as mentioned previously in the implementation of NABC, can also be used to catch detector control attacks. With this measure, in case of normal input (i.e., quantum signals from Alice), the detections in D0 and D1 should no longer bear any relation with the pump phase modulation. In other words, the detections become totally random. Eve's multiphoton states -- replacing the quantum signals during detector control attacks -- would likewise become useless. Even more, based on the detection statistics collected during such checks, Bob would be able to observe a divergent pattern, such as an increased double click rate, thus learning about Eve's attack. 
\subsection{Wavelength-dependent behavior/Trojan-horse \emph{like} attacks\label{protcn:wdbntha}}
Thanks to the various properties of the UCP-QKD receiver, vulnerabilities arising due to the wavelength-dependent behavior of optical components such as fiber couplers and isolators~\cite{Li2011,Jain2015} can be easily prevented. Eve cannot successfully launch an intercept and resend attack at two different wavelengths to exploit the starkly-different splitting ratios of a fiber coupler~\cite{Li2011} since the SPDs are well-isolated from her injected pulses. Traditional Trojan-horse attacks~\cite{Vakhitov2001,Gisin2006,Jain2014}, where Eve sends in a bright pulse into the QKD system and measures back-reflections that may reveal information about the basis choice, can also be thwarted\footnote{In Fig.~\ref{fSysDiag}, the basis choice is implemented by a phase modulator (PM) in the pump arm. An optical path that allows reflection(s), carrying an imprint of the modulation, to propagate out of the system at least physically exists if we assume a fiber-optical wavelength division multiplexer (WDM) instead of the free-space dichroic mirror (DM).}. There are multiple reasons that Eve cannot ascertain the secret basis choice of Bob: the phase of the pump is nonlinearly transferred onto the signal pulses and probing at $\lambda_{\rm pump}$ is prevented via LWI. The only wavelength not protected by LWI is the signal wavelength, and this wavelength is not impacted by the pump phase. 

In general, mechanisms that involve photons carrying useful information inadvertently leaking out of the QKD system without requiring any active injection from Eve are also possible. For instance, in what may be coined as \emph{self} Trojan-horse attack, pump photons carrying information about the state of the PM may incur a back-reflection, say from the waveguide interface, and due to finite extinction/crosstalk of the DM/WDM (used as such for combining the signal and pump; see Fig.~\ref{fSysDiag}), some photons may eventually trickle out of the receiver into the hands of Eve. Alternatively, the flash induced in the single photon avalanche diode during breakdown may reveal information about the clicked detector to an eavesdropper collecting such photons on the quantum channel~\cite{Kurtsiefer2001}. The vulnerabilities due to such mechanisms can also be drastically reduced by a sufficiently-large LWI. 
\subsection{Laser damage attack\label{protcn:dma}}
Attacks aiming to permanently damaging detectors by impinging extremely high CW power~\cite{Bugge2014} are greatly mitigated by the UCP-QKD receiver. We expect that the maximum power injection levels cannot exceed a few Watts due to the fiber fuse effect. Assuming the ability to withstand $\leq 10\,$W input optical power, a large attenuation of $80\,$dB at $\lambda_{\rm pump}$ (as discussed in the previous section) would relegate the power reaching the detectors to a few $\mu$Ws, which can be considered a safe operating condition with respect to permanent damage.

Eve is confined to attack with high powers around $\lambda_{\rm sig}$ due to the LWI property. But even at this wavelength, the PLM property severely restricts the amount of power that eventually reaches the SPDs. To explain via a numerical example, assume the pump with a peak power $P_{out} \approx 100\,$mW and a $10\%$ duty cycle. Even with a unity efficiency conversion of the signal, the average power of the upconverted beam would still be $\lesssim 20\,$mW (factor of 2 from the fact that $\lambda_{\rm sum} \approx 2 \lambda_{\rm pump}$). This power is more than an order of magnitude below what Eve requires to inflict damage on Bob's detectors~\cite{Bugge2014}. Furthermore, through the monitoring system at the exit of the waveguide; see Fig.~\ref{nabcNplm}, Bob would observe not only a high depletion of the pump pulses but also a large response at the signal wavelength. This would clearly result in raising an alarm. 
\subsection{Choice of single photon detectors\label{protcn:chspads}}
The choice of single photon detectors can also impact the security of the QKD system against a variety of hacking attacks. As mentioned earlier, we consider Si-based single photon avalanche diodes (SPADs) that are actively quenched, i.e. operate in the free-running or `ungated' mode. This choice is firstly reasonable because gated mode does not provide any significant improvement in the noise performance of Si-based single photon detectors, unlike InGaAs based SPADs that generally suffer from high thermal dark counts and afterpulsing if left ungated. Secondly, the lack of a detection gate already reduces the chances of many types of detector control attacks. For instance, the (temporal) detection efficiency mismatch loophole~\cite{Makarov2006} that may be exploited via the time-shift attack~\cite{Zhao2008} or the faked-state attack~\cite{Jain2011} can be avoided. This feature also naturally precludes Eve from gaining remote access to the SPADs by attacking \emph{after} the gate~\cite{Wiechers2011}, or at the \emph{falling} edge of the electrical gate\footnote{The detector is in linear mode after the gate and displays superlinear characteristics at the falling edge, i.e. when it is making a transition from the Geiger mode to the linear mode.}~\cite{LydersenJain2011}. Although a UCP-QKD receiver equipped with gated SPADs may resist most attacks, the security constraints are bound to be tighter. Hence, we recommend the use of free-running SPADs. 
\section{Overall performance and practical aspects\label{ovprac}}
In the previous sections, we have described how the UCP-QKD receiver can defeat a plethora of quantum hacking attacks. 
\begin{table}[t!]
\begin{center}
    \begin{tabular}{ |  p{5cm} | l | l | }
    \hline
    \textbf{Parameter name and description} & \textbf{Symbol} & \textbf{Value range (units)} \\ \hline
    Temporal width, signal pulse & $\tau_{\rm sig}$ & $80-100\,$ps \\ \hline
    Temporal width, pump pulse & $\tau_{\rm pump}$ & $100-120\,$ps \\ \hline
    System repetition rate & $f_{R}$ & $1.0\,$GHz \\ \hline
    Center wavelength, signal & $\lambda_{\rm sig}$ & $1550\pm 20\,$nm \\ \hline
    Center wavelength, pump  & $\lambda_{\rm pump}$ & $1830\pm 20\,$nm \\ \hline
    Center wavelength, sum & $\lambda_{\rm sum}$ & $830\pm 10\,$nm \\ \hline		
    Bandwidth, signal BPF & $\Delta_{\rm sig}$ & $10-20\,$nm \\ \hline
		Bandwidth, sum frequency BPF & $\Delta_{\rm sum}$ & $2-5\,$nm \\ \hline
		Temporal separation between time bins corresponding to bits ``0'' and ``1'' & $\delta T$ & $300-400\,$ps \\ \hline
    \end{tabular}
\end{center}
\caption{Values of various significant parameters in the upconversion based quantum key distribution (QKD) receiver shown in Fig.~\ref{fSysDiag}. \label{tTyplVals}}
\end{table}
Table~\ref{tTyplVals} lists the range of values of different parameters that have been already introduced in the last pages. This represents one possible design of the system. Note that many design alternatives are possible, such as using $2$ micron pump light to reduce noise even further, provided they are engineered to preserve the basic properties of a UCP-QKD receiver, described in subsection~\ref{ucdfQKDrxr:bb}.

The upconversion process can be highly efficient: values of $\eta_c > 80\%$ for the conversion efficiency, which includes the optical transmission and the in- and out- coupling losses through the waveguide, are viable~\cite{Pelc2011}. High-performance detectors exhibiting large quantum efficiencies $\eta_d \approx 80\%$ at the upconverted wavelengths, low intrinsic noise characteristics $D \approx 10-100\,$cps, and fast counting rates $R \gtrsim 10\,$Mcps are commercially available. The probability of successfully upconverting and detecting a quantum signal pulse at $\lambda_{\rm sig}$ is then given by the overall efficiency $\eta_{ov} = \eta_c \eta_d \eta_f$; values in the range of $30-40\%$ can be envisaged easily. Here $\eta_f \approx 50\%$ denotes the cumulative transmission of the various optical filters used for implementing LWI; see Fig.~\ref{fSysDiag}. 

Based on these numbers, sifted key rates of the order of $100\,$Kbps for $\sim 100\,$km channel lengths would be feasible with the first generation receivers. Also, by adopting measures such as long wavelength pump and strict filtering, the overall noise in the upconversion process can also be kept sufficiently low; assuming $D_{ov} \leq 1\,$Kcps, the contribution to the quantum bit error rate would be under $1\%$. We note that for ease of implementation one can also use a pump source very close to the signal (e.g. $10\,$nm~\cite{Kowligy2014}), where high-quality modulators and amplifiers are readily available. However, this would increase noise and may make the implementation of LWI more difficult. 

As mentioned in the introduction, UCP-QKD is similar to MDI-QKD in the sense that both address vulnerabilities in the photon detection apparatus of realistic QKD implementations. Also, both offer realistic implementations with off-the-shelf components; most significantly, both allow the transmitter to use weak coherent states, typically realized by attenuated laser pulses, which are the most practical quantum signals for QKD. While only MDI-QKD claims a rigorous security proof against quantum hacking of the detectors, UCP-QKD does not share some of MDI-QKD's stringent requirements that make the latter more difficult to implement and lower the resulting key rate. For instance MDI-QKD implementations require indistinguishability of the weak coherent states generated by both Alice and Bob and they also intrinsically suffer from a reduced key rate due to `low gain' of the single photon states~\cite{Liu2013}. Thus, although the two methods have fundamental differences which make a direct comparison difficult, roughly speaking, while MDI-QKD offers a more secure but lower performing system, the goal of UCP-QKD is improved security without a substantial reduction in performance or practicality. 
\section{Conclusion\label{conc}}
In conclusion, we have detailed the design and operation of a novel upconversion protected (UCP) receiver for the purpose of doing secure quantum key distribution (QKD). We have specifically shown how Bob can use this UCP-QKD receiver to protect himself against various quantum hacking attacks, launched by an eavesdropper Eve from the quantum channel. This is possible because of properties intrinsic to the upconversion process that can restrict the degrees of freedom which Eve might have otherwise exploited to take advantage of the imperfections of the detection apparatus. Eve's manipulations are thus either rendered futile or can be suitably monitored to raise alarms and abort the QKD protocol. The security features can even be combined with other known measures to counter attacks, such as the real time monitoring of the detector parameters~\cite{Silva2012}. Such UCP-QKD receivers therefore greatly enhance the security of practical QKD systems. With rapid advances in integrated photonic technologies, on-chip implementations of the waveguide and filtering system can minimize the optical losses even further, thereby increasing the key rates obtainable with such receivers. 
\begin{acknowledgements}
This research was supported in part by the DARPA Quiness program (Grant number: W31P4Q-13-1-0004).
\end{acknowledgements}
\bibliographystyle{spphys}       
\bibliography{library}

\begin{thebibliography}{10}
\providecommand{\url}[1]{{#1}}
\providecommand{\urlprefix}{URL }
\expandafter\ifx\csname urlstyle\endcsname\relax
  \providecommand{\doi}[1]{DOI \discretionary{}{}{}#1}\else
  \providecommand{\doi}{DOI \discretionary{}{}{}\begingroup
  \urlstyle{rm}\Url}\fi

\bibitem{Midwinter1967}
J.~Midwinter, J.~Warner, Journal of Applied Physics \textbf{38}(2), 519 (1967)

\bibitem{Tucker1969}
J.~Tucker, D.F. Walls, Annals of Physics \textbf{52}(1), 1  (1969)

\bibitem{Kumar1990}
P.~Kumar, Opt. Lett. \textbf{15}(24), 1476 (1990)

\bibitem{Huang1992}
J.~Huang, P.~Kumar, Phys. Rev. Lett. \textbf{68}, 2153 (1992)

\bibitem{Kim2001}
Y.H. Kim, S.P. Kulik, Y.~Shih, Phys. Rev. Lett. \textbf{86}, 1370 (2001)

\bibitem{Vandevender2004}
A.P. Vandevender, P.G. Kwiat, Journal of Modern Optics \textbf{51}(9-10), 1433
  (2004)

\bibitem{Diamanti2005}
E.~Diamanti, H.~Takesue, T.~Honjo, K.~Inoue, Y.~Yamamoto, Phys. Rev. A
  \textbf{72}, 052311 (2005)

\bibitem{MaL2012}
L.~Ma, O.~Slattery, X.~Tang, Physics Reports \textbf{521}(2), 69  (2012)

\bibitem{Thew2006}
R.T. Thew, S.~Tanzilli, L.~Krainer, S.C. Zeller, A.~Rochas, I.~Rech, S.~Cova,
  H.~Zbinden, N.~Gisin, New Journal of Physics \textbf{8}(3), 32 (2006)

\bibitem{Pelc2011}
J.S. Pelc, L.~Ma, C.R. Phillips, Q.~Zhang, C.~Langrock, O.~Slattery, X.~Tang,
  M.M. Fejer, Opt. Express \textbf{19}(22), 21445 (2011)

\bibitem{Pelc2012}
J.S. Pelc, Q.~Zhang, C.R. Phillips, L.~Yu, Y.~Yamamoto, M.M. Fejer, Opt. Lett.
  \textbf{37}(4), 476 (2012)

\bibitem{Eckstein2011}
A.~Eckstein, B.~Brecht, C.~Silberhorn, Opt. Express \textbf{19}(15), 13770
  (2011)

\bibitem{Huang2013}
Y.P. Huang, P.~Kumar, Opt. Lett. \textbf{38}(4), 468 (2013)

\bibitem{Kowligy2014}
A.S. Kowligy, P.~Manurkar, N.V. Corzo, V.G. Velev, M.~Silver, R.P. Scott,
  S.J.B. Yoo, P.~Kumar, G.S. Kanter, Y.P. Huang, Opt. Express \textbf{22}(23),
  27942 (2014)

\bibitem{Tanzilli2005}
S.~Tanzilli, W.~Tittel, M.~Halder, O.~Alibart, P.~Baldi, N.~Gisin, H.~Zbinden,
  Nature \textbf{437}(7055), 116 (2005)

\bibitem{Gisin2007}
N.~Gisin, R.~Thew, Nature Photonics \textbf{1}, 165 (2007)

\bibitem{Kimble2008}
H.J. Kimble, Nature \textbf{453}(7198), 1023 (2008)

\bibitem{BB84}
C.H. Bennett, G.~Brassard, in \emph{Proceedings of IEEE International
  Conference on Computers Systems and Signal Processing} (Bangalore India,
  1984), pp. 175--179

\bibitem{Gisin2002}
N.~Gisin, G.~Ribordy, W.~Tittel, H.~Zbinden, Reviews of Modern Physics
  \textbf{74}(1), 145 (2002)

\bibitem{Scarani2009a}
V.~Scarani, H.~Bechmann-Pasquinucci, N.~Cerf, M.~Du\v{s}ek, N.~L\"{u}tkenhaus,
  M.~Peev, Reviews of Modern Physics \textbf{81}(3), 1301 (2009)

\bibitem{Inoue2002}
K.~Inoue, E.~Waks, Y.~Yamamoto, Physical Review Letters \textbf{89}(3), 37902
  (2002)

\bibitem{Zhang2009}
Q.~Zhang, H.~Takesue, T.~Honjo, K.~Wen, T.~Hirohata, M.~Suyama, Y.~Takiguchi,
  H.~Kamada, Y.~Tokura, O.~Tadanaga, Y.~Nishida, M.~Asobe, Y.~Yamamoto, New
  Journal of Physics \textbf{11}(4), 045010 (2009)

\bibitem{Lo2012}
H.K. Lo, M.~Curty, B.~Qi, Physical Review Letters \textbf{108}(13), 130503
  (2012)

\bibitem{Liu2013}
Y.~Liu, T.Y. Chen, L.J. Wang, H.~Liang, G.L. Shentu, J.~Wang, K.~Cui, H.L. Yin,
  N.L. Liu, L.~Li, X.~Ma, J.S. Pelc, M.M. Fejer, C.Z. Peng, Q.~Zhang, J.W. Pan,
  Physical Review Letters \textbf{111}(13), 130502 (2013)

\bibitem{Xu2007}
H.~Xu, L.~Ma, A.~Mink, B.~Hershman, X.~Tang, Opt. Express \textbf{15}(12), 7247
  (2007)

\bibitem{Scarani2009}
V.~Scarani, C.~Kurtsiefer, arXiv:0906.4547

\bibitem{Lo2014}
H.K. Lo, M.~Curty, K.~Tamaki, Nature Photonics \textbf{8}(8), 595 (2014)

\bibitem{VadimWebsite}
{Website, Vadim Makarov}.
\newblock \url{http://www.vad1.com}

\bibitem{Wiechers2011}
C.~Wiechers, L.~Lydersen, C.~Wittmann, D.~Elser, J.~Skaar, C.~Marquardt,
  V.~Makarov, G.~Leuchs, New Journal of Physics \textbf{13}(1), 013043 (2011)

\bibitem{Lydersen2010a}
L.~Lydersen, C.~Wiechers, C.~Wittmann, D.~Elser, J.~Skaar, V.~Makarov, Optics
  Express \textbf{18}(26), 27938 (2010)

\bibitem{Gerhardt2011}
I.~Gerhardt, Q.~Liu, A.~Lamas-Linares, J.~Skaar, C.~Kurtsiefer, V.~Makarov,
  Nature Communications \textbf{2}, 349 (2011)

\bibitem{Sauge2011}
S.~Sauge, L.~Lydersen, A.~Anisimov, J.~Skaar, V.~Makarov, Opt. Express
  \textbf{19}(23), 23590 (2011)

\bibitem{Jain2011}
N.~Jain, C.~Wittmann, L.~Lydersen, C.~Wiechers, D.~Elser, C.~Marquardt,
  V.~Makarov, G.~Leuchs, Physical Review Letters \textbf{107}(11), 5 (2011)

\bibitem{LydersenJain2011}
L.~Lydersen, N.~Jain, C.~Wittmann, O.~Maroy, J.~Skaar, C.~Marquardt,
  V.~Makarov, G.~Leuchs, Physical Review A \textbf{84}(3), 032320 (2011)

\bibitem{Weier2011}
H.~Weier, H.~Krauss, M.~Rau, M.~F\"{u}rst, S.~Nauerth, H.~Weinfurter, New
  Journal of Physics \textbf{13}(7), 073024 (2011)

\bibitem{Bugge2014}
A.N. Bugge, S.~Sauge, A.M.M. Ghazali, J.~Skaar, L.~Lydersen, V.~Makarov,
  Physical Review Letters \textbf{112}(7), 70503 (2014)

\bibitem{Li2011}
H.W. Li, S.~Wang, J.Z. Huang, W.~Chen, Z.Q. Yin, F.Y. Li, Z.~Zhou, D.~Liu,
  Y.~Zhang, G.C. Guo, W.S. Bao, Z.F. Han, Physical Review A \textbf{84}(6),
  062308 (2011)

\bibitem{Jain2014}
N.~Jain, E.~Anisimova, I.~Khan, V.~Makarov, C.~Marquardt, G.~Leuchs, New
  Journal of Physics \textbf{16}(12), 123030 (2014)

\bibitem{Jain2015}
N.~Jain, B.~Stiller, I.~Khan, V.~Makarov, C.~Marquardt, G.~Leuchs, IEEE Journal
  of Selected Topics in Quantum Electronics \textbf{21}(3), 1 (2015)

\bibitem{Kanter2015}
G.S. Kanter, in \emph{CLEO: 2015} (Optical Society of America, 2015), p. JW2A.7

\bibitem{Silva2012}
T.~{Ferreira da Silva}, G.B. Xavier, G.P. Tempor\~{a}o, J.P. von~der Weid,
  Optics Express \textbf{20}(17), 18911 (2012)

\bibitem{Lucamarini2015}
M.~Lucamarini, I.~Choi, M.B. Ward, J.F. Dynes, Z.L. Yuan, A.J. Shields,
  Physical Review X \textbf{5}(3), 031030 (2015)

\bibitem{AchillesHeelQKD}
C.H. Bennett.
\newblock {Let Eve do the heavy lifting, while John and Won-Young keep her
  honest.}
\newblock \url{http://dabacon.org/pontiff/?p=5340}

\bibitem{Dong2008}
H.~Dong, H.~Pan, Y.~Li, E.~Wu, H.~Zeng, Applied Physics Letters \textbf{93}(7),
  071101 (2008)

\bibitem{Itzler2011}
M.A. Itzler, X.~Jiang, M.~Entwistle, K.~Slomkowski, A.~Tosi, F.~Acerbi,
  F.~Zappa, S.~Cova, Journal of Modern Optics \textbf{58}(3-4), 174 (2011)

\bibitem{Bennett92c}
C.H. Bennett, Physical Review Letters \textbf{68}(21), 3121 (1992)

\bibitem{Hughes2000}
R.J. Hughes, G.L. Morgan, C.G. Peterson, Journal of Modern Optics
  \textbf{47}(2-3), 533 (2000)

\bibitem{Ekert1992}
A.~Ekert, J.G. Rarity, P.~Tapster, G.~{Massimo Palma}, Physical Review Letters
  \textbf{69}(9), 1293 (1992)

\bibitem{Marcikic2004}
I.~Marcikic, H.~de~Riedmatten, W.~Tittel, H.~Zbinden, M.~Legr\'e, N.~Gisin,
  Phys. Rev. Lett. \textbf{93}, 180502 (2004)

\bibitem{Yuan2005a}
Z.~Yuan, A.~Shields, Optics Express \textbf{13}(2), 660 (2005)

\bibitem{Boyd2008}
R.W. Boyd, \emph{Nonlinear Optics, Third Edition}, 3rd edn. (Academic Press,
  2008)

\bibitem{Gisin2006}
N.~Gisin, S.~Fasel, B.~Kraus, H.~Zbinden, G.~Ribordy, Physical Review A
  \textbf{73}(2), 1 (2006)

\bibitem{ThorlabsDMSumPumpSep}
{Dichroic Mirrors/Beamsplitters: 1180 nm Cutoff Wavelength}.
\newblock \url{www.thorlabs.com}

\bibitem{OmegaDMSumPumpSep}
{Dichroic filter 1400BS}.
\newblock \url{www.omegafilters.com}

\bibitem{Tanzilli2012}
S.~Tanzilli, A.~Martin, F.~Kaiser, M.~De~Micheli, O.~Alibart, D.~Ostrowsky,
  Laser \& Photonics Reviews \textbf{6}(1), 115 (2012)

\bibitem{Harris2014}
N.C. Harris, D.~Grassani, A.~Simbula, M.~Pant, M.~Galli, T.~Baehr-Jones,
  M.~Hochberg, D.~Englund, D.~Bajoni, C.~Galland, Phys. Rev. X \textbf{4},
  041047 (2014)

\bibitem{opnetiWDM}
{1x2 1310/1550nm High Isolation Filter WDM}.
\newblock \url{www.opneti.com}

\bibitem{Edmund85891}
{1550nm CWL, 12.5mm Dia. Hard Coated OD 4 50nm Bandpass Filter}.
\newblock \url{www.edmundoptics.com}

\bibitem{SemrockLL01830}
{830 nm MaxLine laser clean-up filter}.
\newblock \url{www.semrock.com}

\bibitem{Lydersen2010}
L.~Lydersen, C.~Wiechers, C.~Wittmann, D.~Elser, J.~Skaar, V.~Makarov, Nature
  Photonics pp. 1--10 (2010)

\bibitem{Zhao2008}
Y.~Zhao, C.H. Fung, B.~Qi, C.~Chen, H.K. Lo, Physical Review A \textbf{78}(4),
  042333 (2008)

\bibitem{Vakhitov2001}
A.~Vakhitov, V.~Makarov, D.~Hjelme, Journal of Modern Optics \textbf{48}(13),
  2023 (2001)

\bibitem{Kurtsiefer2001}
C.~Kurtsiefer, P.~Zarda, S.~Mayer, H.~Weinfurter, Journal of Modern Optics
  \textbf{48}(13), 2039 (2001)

\bibitem{Makarov2006}
V.~Makarov, A.~Anisimov, J.~Skaar, Physical Review A \textbf{74}(2), 1 (2006)

\end{thebibliography}

\end{document}